\documentclass[aps,prb,twocolumn,groupedaddress]{revtex4}

\usepackage{amsmath}
\usepackage{graphicx}
\begin{document}


\def\be{\begin{equation}}
\def\ee{\end{equation}}
\def\beq{\begin{align}}
\def\eeq{\end{align}}
\def\oc{\omega_{\rm c}} 
\def\lb{l_{\rm B}}  
\def\kb{k_{\rm B}}      
\def\me{m_{\rm e}}      
\def\D{{\rm d}}         
\def\I{{\rm i}}         
\def\nel{n_{\rm e}}
\def\Nel{N_{\rm e}}
\def\kf{{k_{\rm F}}}
\def\E{{\rm e}}

\title{Local Density of States around an Impurity
in a Strong Magnetic Field. \\
I. a Two-Dimensional System with Parabolic Dispersion}


\author{Daijiro \textsc{Yoshioka}}
\affiliation{Department of Basic Science, The University of Tokyo \\
Komaba, Meguro-ku, Tokyo, 153-8902, Japan\\}


\date{October 17, 2006}

\begin{abstract}
Bound states around an impurity are investigated for a two dimensional electron system in a strong magnetic field.
Long-range Coulomb potential and related potentials are considered.
Schr\"odinger equation is solved numerically to obtain the bound states.
The energy and wave function of these bound states are indirectly observed by the scanning tunneling spectroscopy as local density of states (LDOS).
Theoretically obtained LDOS is compared with experiment.
Reasonable agreement is obtained.
\end{abstract}


\maketitle



%



\section{Introduction} 

In recent experiments, Matsui et al.\cite{matsui} and Niimi et al.\cite{niimi} succeeded to observe local density of states at a surface of graphite by scanning tunneling spectroscopy (STS).
In a strong magnetic field at low temperature, they observed a central peak and ring-like structure around impurities.
Graphite is a three dimensional material; it is a semimetal having both electron band and hole band.
However, HOPG (highly oriented pyrolytic graphite) that is used in the experiments looks like two-dimensional when observed by the transport and STS; the two dimensional Landau levels are observed, and quantum Hall like plateau is observed.\cite{kopelevich}
The origin of the two-dimensionality is attributed to abundance of stacking faults in this type of graphite.
In this paper we examine to what extent the observed structure can be understood by an ideal two-dimensional model, and clarify the meaning of the structure observed in these experiments.

Although graphite has both electron band and hole band, here we consider simple 2-dimensional electron system with parabolic dispersion.
We place a positively charged impurity at the origin of the two-dimensional plane, or some distance $d$ separated from the origin in the direction perpendicular to the 2-d plane, taking into account the possibility that the impurity is situated in the bulk of graphite.
In this model, electronic states around the origin in a strong magnetic field are investigated.

The angular momentum around the origin is a good quantum number, and the bound states of definite angular momentum are formed below each Landau levels.
Which value of the angular momentum gives the lowest energy state for each Landau level depends on the distance $d$.
When $d$ is less than the order of the magnetic length, the states with angular momentum zero are the lowest energy states for each Landau level.
If $d$ is much smaller than the magnetic length, the wave function of these states has a sharp peak at the origin, and the energy is separated from other states.
In such a case the local density of states (LDOS) at the energy between the Landau levels shows a sharp central peak surrounded by ring-like structure.
This shape of LDOS resembles experimental results.
On the other hand, when $d$ becomes large, such structure as observed experimentally is lost.
Thus, we claim that the observation of the LDOS by STS can give information on the possible structure of the impurity potential.

This investigation has been begun as collaboration with Fukuyama's group for the explanation of their experimental data.
Part of the present results are included in Niimi et al's letter.\cite{niimi}
This paper gives general treatment of the problem and details of the calculation.

Recently, truly two-dimensional graphite is realized, which is known as graphene.\cite{novoselov,novoselov2,zhang}
In this case the energy spectrum is not parabolic, but linear, and carriers there are sometimes called massless Dirac fermions.
How the LDOS around an impurity in this material shows up is also an interesting problem.
Theoretical investigation for this case is given in a forth coming paper.

This paper is organized as follows.
In \S 2 potential used in this calculation and the method are explained.
In \S 3 we notice that the wave function and energy of positive angular momentum state is known from those of the negative angular momentum state in the presence of isotropic potential of arbitrary shape.
The concrete energy and wave function in the presence of impurity potential are obtained in \S 4 and 5.
We calculate local density of states in \S 6 using the results of energy and wave functions.
Finally in \S 7 comparison with experiment is done.

\section{Model and Method}

We consider a model of two-dimensional electron system in a magnetic field with a positively charged impurity at the origin.
The Coulomb interaction between electrons is neglected, so we consider single electron states in this paper.
Being a single electron problem, the spin freedom is also neglected.
Then the Hamiltonian is 
\be
H = H_0 + V(\vec{r})\,,
\ee
where
\be
H_0=\frac{1}{2\me}(\vec{p} -e \vec{A})^2 \,,
\ee
is the unperturbed Hamiltonian, and $V(\vec{r})$ is the impurity potential.
Here, $\me$ is electron's band mass, and $e<0$ is its charge.
The momentum $\vec{p}=(p_x,p_y)$, the coordinate $\vec{r}=(x,y)$, and the vector potential $\vec{A}=(A_x,A_y)$ are two-dimensional vectors.
A uniform magnetic field $B$ is applied perpendicular to the 2-d plane, so the vector potential is given as $\vec{A}=(-By/2,Bx/2)$ in the symmetric gauge.

Considering a possibility that the impurity is situated at a distance $d$ from the 2-d plane, we investigate a system where the impurity potential is given as
\be
V(\vec{r}) = - \frac{e^2}{4\pi\epsilon \sqrt{r^2+d^2}} \,,
\ee
where $\epsilon$ is the dielectric constant.
This potential includes two limiting cases; one is pure Coulomb case, $d=0$, and the other is a case where $d$ is large enough that the potential around the origin can be approximated by a harmonic potential:
\be
V(\vec{r}) =\frac{1}{2}\me\omega_0^2r^2 + {\rm const.}\,,
\ee
where $\omega_0^2=e^2/4\pi\epsilon d^3\me$, and $\omega_0$ gives frequency of the harmonic oscillation of an electron around the origin in the absence of the magnetic field.

The eigenstates of this Hamiltonian in the limit of large $d$, the harmonic potential case, has been analytically obtained and well known,\cite{fock}  but those with general values of $d$ are not.
Here, we solve this problem numerically.
Namely, since we know the eigenstates of the unperturbed Hamiltonian, the Landau states, we calculate the matrix elements of the potential, and diagonalize the Hamiltonian with the basis of Landau states.

The procedure is straight forward.
However, before giving the results, we summarize the eigenstates of the unperturbed Hamiltonian $H_0$, and also the eigenstates with the harmonic potential.
One way to diagonalize $H_0$ is to use a set of canonical variables, the dynamical momentum $\vec{\pi}$ and the center coordinate $\vec{R}=(X,Y)$ instead of canonical momentum $\vec{p}=(p_x,p_y)$ and coordinate $\vec{r}=(x,y)$.
They are defined as
\begin{equation}
\vec{\pi} = \vec{p}-e\vec{A}
\end{equation}
and
\begin{align}
X&=x-\frac{\lb^2}{\hbar}\pi_y \,,\\
Y&=y+\frac{\lb^2}{\hbar}\pi_x \,,
\end{align}
where
$\lb \equiv \sqrt{\hbar/|e|B}$ is the magnetic length.
These variables satisfy commutation relations,
\be
[\pi_x,\pi_y]=-\I \frac{\hbar^2}{\lb^2}\,,
\ee
and
\be
[X,Y]=\I \lb^2\,.
\ee
In terms of these variables, the Hamiltonian $H_0$ and the angular momentum $L_z$ are written in quadratic forms,
\be
H_0 = \frac{1}{2\me} \left(\pi_x^2+\pi_y^2\right) \,,
\ee
and 
\be
L_z = - \frac{\hbar}{2\lb^2}\left(X^2+Y^2\right) + \frac{\lb^2}{2\hbar}\left(\pi_x^2+\pi_y^2\right) \,.
\ee
Then by introducing ladder operators for the Landau level,
\be
a^\dagger = \frac{\lb}{\sqrt{2}\hbar}\left( \pi_x+\I \pi_y\right) \,,
\ee
\be
a= \frac{\lb}{\sqrt{2}\hbar}\left( \pi_x - \I \pi_y\right) \,,
\ee
and those for the angular momentum,
\be
b^\dagger = \frac{1}{\sqrt{2}\lb}\left(X - \I Y \right) \,,
\ee
\be
b = \frac{1}{\sqrt{2}\lb}\left(X + \I Y \right) \,,
\ee
we rewrite $H_0$ and $L_z$ as follows
\be
H_0 = \hbar\oc \left(a^\dagger a + \frac{1}{2} \right) \,,
\ee
and
\be
L_z = \hbar \left(a^\dagger a -b^\dagger b\right) \,.
\ee
Here $\oc=|e|B/\me$ is the cyclotron frequency.
Simultaneous eigenstates of $a^\dagger a$ and $b^\dagger b$, 
$|n,m\rangle$, which satisfies
\be
a^\dagger a |n,m\rangle = n |n,m\rangle \,,
\ee
\be
b^\dagger b |n,m\rangle = m |n,m\rangle \,,
\ee
is also the eigenstates of $H_0$ and $L_z$, and eigenvalues are given as
$E_{n} = (n+1/2)\hbar\oc$ and $l_z\hbar=(n-m)\hbar$, respectively with $n$ and $m$ being non-negative integers.

The eigenstates in the presence of the harmonic potential $V(r)= (1/2)\me\omega_0^2r^2$ are obtained by our using the ladder operators $\tilde a^\dagger$, $\tilde a$, $\tilde b^\dagger$, and $\tilde b$ at larger magnetic field,
\be
\tilde B \equiv \sqrt{1+ 4\frac{\omega_0^2}{\oc^2}}\, B \,.
\ee
In terms of them, $H$ and $L_z$ are expressed as follows
\beq
H &= H_0 + \frac{1}{2}\me\omega_0^2r^2 \nonumber\\
&=
\frac{\hbar}{2}\sqrt{\oc^2+4\omega_0^2} \left(\tilde a^\dagger \tilde a + \tilde b^\dagger \tilde b +1\right)
+ \frac{\hbar}{2}\oc\left(\tilde a^\dagger \tilde a - \tilde b^\dagger \tilde b\right)\,,
\end{align}
and
\be
L_z = \hbar \left(\tilde a^\dagger \tilde a - \tilde b^\dagger \tilde b\right)\,.
\ee
Thus, the eigenstate of the ladder operators at $\tilde B$, $|n,m\rangle_{\tilde B}$, is the eigenstates of $H$ and $L_z$ at $B$ with eigenvalues,
\begin{eqnarray}
&&H|n,m\rangle_{\tilde B}= E_{n,m}|n,m\rangle_{\tilde B}\nonumber\\
&&= \left\{ \frac{\hbar}{2}\sqrt{\oc^2+4\omega_0^2} (n+m+1)
+ \frac{\hbar}{2}\oc(n-m)\right\} |n,m\rangle_{\tilde B}\,,\nonumber\\
\end{eqnarray}
and
\be
L_z|n,m\rangle_{\tilde B} = \hbar (n-m)|n,m\rangle_{\tilde B} \,.
\ee
It should be noted that in this case the lowest energy state for each Landau level is the state with $m=0$, or angular momentum $l_z=n$.

In order to obtain the eigenstates of the Hamiltonian in the presence of the Coulomb potential $V(r)$, we express the total Hamiltonian as a matrix using the eigenstates of $H_0$ and $L_z$, and numerically diagonalize it.
For numerical calculation we parameterize the Coulomb potential:
\be
V(r) = \hbar\oc \frac{\alpha}{\sqrt{\xi^2+\delta^2}} \,,
\ee
where $\xi=r/\lb$, $\delta = d/\lb$ and $\alpha = e^2/4\pi\epsilon \hbar\oc\lb$.
Namely, we use $\oc$ and $\lb$ as units of energy and length, respectively.
In the following we use a symbol $\mathcal{E}$ for energy expressed in the units of $\hbar\oc$, namely, $E=\mathcal{E}\hbar\oc$.

Even in the presence of $V(r)$, the angular momentum remains to be a good quantum number.
So the Hamiltonian is block diagonalized for each angular momentum.
For a given $l_z=n-m$, it is expressed as
\begin{widetext}
\beq
\frac{H}{\hbar\oc} &= \sum_n |n,n-l_z\rangle \left(n+\frac{1}{2}\right)\langle n,n-l_z| \nonumber\\
&+ \alpha\sum_{n_1,n_2} |n_1,n_1-l_z\rangle \langle n_1,n_1-l_z| \frac{1}{\sqrt{\xi^2+\delta^2}} |n_2,n_2-l_z\rangle \langle n_2,n_2-l_z| \,.
\end{align}
For the case of $d/\lb =\delta =0$, the matrix elements are calculated analytically:
\begin{align}
\langle n_1,n_1-l_z| &\displaystyle{\frac{1}{\xi}} |n_2,n_2-l_z\rangle =
\frac{1}{\sqrt{2}}\left( \frac{N_1!N_2!}{(N_1+|l_z|)!(N_2+|l_z|)!}  \right)^{1/2}\nonumber\\
&\times \sum_{j=0}^{{\rm min}({N_1,N_2})} \frac{\Gamma(j+|l_z|+\frac{1}{2})}{2^{2N_1+2N_2-4j-2}j!}
\left(\begin{array}{c}2N_1-2j-1\\N_1-j \end{array} \right)
\left(\begin{array}{c}2N_2-2j-1\\N_2-j \end{array} \right)\,,
\end{align}
\end{widetext}
where 
\be
N_{1,2} \equiv n_{1,2} - \frac{1}{2}(l_z+|l_z|) \,,
\ee
and $\Gamma$ is the gamma function.
We label the eigenstates by $n$ and $m$ so that they evolve continuously as $\alpha$ increases.

\section{Relation between Positive and Negative Angular Momenta}

In our system with isotropic impurity potential, the angular momentum is conserved.
When the direction of the magnetic field is in the positive $z$-direction, which is the choice in this paper, a state in the lowest Landau level, $n=0$, has only zero or negative angular momentum.
In the higher Landau levels ($n > 0$), most states have negative $l_z$, but those with $0 \le m <n$ have positive angular momentum.
The energy and wave function of these states with a positive angular momentum $l_z$ are known once we know the eigenstate with angular momentum $-l_z$ in the lower Landau levels.
We explain this relation in this section.

The Hamiltonian in the polar coordinate $(r,\theta)$ is 
\begin{align}
H=&-\frac{\hbar^2}{2\me}\left[\frac{1}{r}\,\frac{\partial}{\partial r}\left(r\,\frac{\partial}{\partial r}  \right)
+ \frac{1}{r^2}\frac{\partial^2}{\partial \theta^2} \right] \nonumber\\ 
&+ \I \frac{1}{2}\hbar\oc \frac{\partial}{\partial \theta}
+\frac{1}{2}\me\oc^2 r^2 +V(r)\,.
\end{align}
We write the wave function as
\be
\psi(r,\theta) = \E^{\I l_z\theta} R(r) \,.
\ee
Then the Schr\"odinger equation for $R(r)$ with eigenenergy $E$ is
\begin{widetext}
\be
-\frac{\hbar^2}{2\me}\left(R''+\frac{1}{r}R' -\frac{l_z^2}{r^2}R \right) + \left[ \frac{1}{2}\me \oc^2 r^2 + V(r) \right] R = \left(E-\frac{1}{2}\hbar\oc l_z \right)R\,.
\ee
\end{widetext}
In the left hand side of this equation only $l_z^2$ appears.
Therefore, we can see that we obtain the same $R(r)$ for both positive and negative angular momenta.
For the same $R$, the eigenenergies are different, since the right hand side has bare $l_z$.
If we write the energy at positive $l_z$ as $E_>$ and that at negative angular momentum $l_z'=-l_z$ as $E_<$, these energies are related with the following equation:
\be
E_> - \frac{1}{2}\hbar\oc l_z = E_< - \frac{1}{2}\hbar\oc l_z' =E_< + \frac{1}{2}\hbar\oc l_z\,,
\ee
Namely, the energy of a state with a positive angular momentum $l_z$ is higher than the negative angular momentum counterpart by $l_z\hbar\oc$: $E_>=E_<+l_z\hbar\oc$.
In terms of the labelling of the eigenstates with $n$ and $m$, we can say that the wave functions are related as
\be
\langle\vec{r}|n,m\rangle \E^{\I m\theta} = \langle\vec{r}|m,n\rangle \E^{\I n\theta} \,,
\ee
and
\be
E_{n,m} = E_{m,n} + (n-m) \hbar\oc \,.
\ee
It is easily checked that the eigenstates in harmonic potential have this property.
Considering the fact that the classical orbits of these states are quite different, this result for the radial part of the wave function is quite noteworthy.

\section{Energy levels}

\begin{figure}[tb]
\begin{center}
\includegraphics[width=7cm]{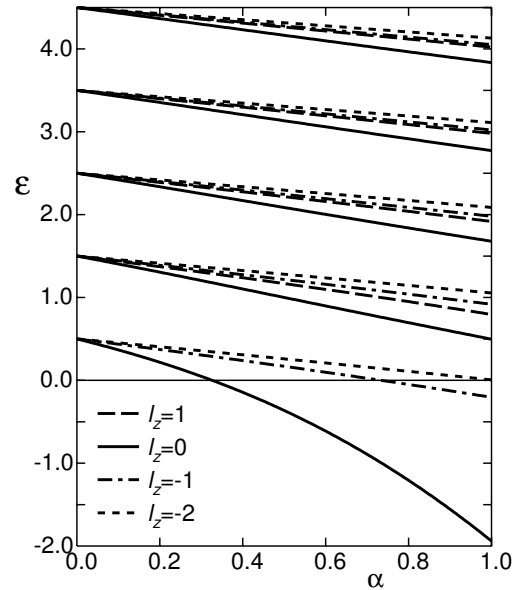}
\end{center}
\caption{Splitting of the Landau levels in the presence of $1/r$ potential.
Energy of the states in the units of $\hbar\oc$, $\mathcal{E}$, is shown as functions of the potential strength $\alpha$.
States with angular momentum $l_z \hbar=(n-m)\hbar = -2\hbar$ to $\hbar$ are shown.}
\label{fig1}
\end{figure}

In the absence of the impurity potential, the energy spectrum consists of Landau levels, and they are degenerate with respect to $m$.
In the presence of the impurity potential, the degeneracy is lifted.
The resultant energy spectrum for $\delta=0$ in the units of $\hbar\oc$, $\mathcal{E}$, is shown in Fig.~\ref{fig1} as functions of the potential strength $\alpha$.
In this figure states with angular momentum $l_z \hbar= (n-m) \hbar=  -2 \hbar$ to $\hbar$ are shown.
As shown in this figure the lowest energy state among a given Landau levels is always the $l_z=0$ state.
This is in contrast to the case of harmonic potential, in which case the lowest energy states are those with $m=0$, or $l_z=n$.
Thus, except for the lowest Landau level, the sequence of the energy levels is different.

\begin{figure}[tb]
\begin{center}
\includegraphics[width=6cm]{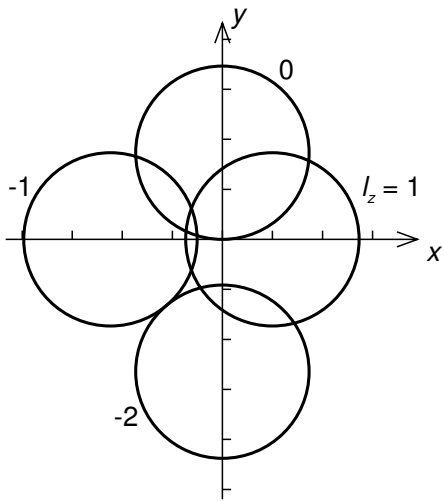}
\end{center}
\caption{Classical cyclotron orbits of an electron which has quantum number $n=1$ and $m=0$ to 3, or $l_z=-2$ to 1.  The values of $l_z$ is written by each orbit.}
\label{fig2}
\end{figure}

The reason for the different behavior is understood when we consider corresponding classical orbits for each $n$ and $m$.
In the absence of the potential, the classical cyclotron orbit is a circle centered at the center coordinate $\vec{R}=(X,Y)$ with radius $r_0$.
The energy and the angular momentum in the classical mechanics are given as
\be
E=\frac{1}{2}m_{\rm e} \oc^2r_0^2 \,,
\ee
and
\be
L_z= \frac{1}{2}|e|B\left(r_0^2 - R^2\right)
\ee
Thus, the Landau quantum number $n$ is related to $r_0^2$, and $m$ is related to $R^2$.
Namely, $r_0^2=(2n+1)\lb^2$ and $R^2=(2m+1)\lb^2$.
The classical orbits for the second lowest Landau level ($n=1$) with $m=0 $ to 3 or $l_z=-2$ to 1 are shown in Fig.~\ref{fig2}, where the positions of the orbital centers are placed at distance $R$ from the origin, the direction angle being increasing by $\pi/2$.
If we consider the effect of the impurity potential perturbatively, the average of the impurity potential along the orbit gives the energy gain from the potential.
Since the classical orbit with $l_z=0$ always passes through the origin, where the $1/r$ potential diverges, it has the lowest energy.
On the other hand, the orbit with $m=0$ has orbit center nearest to the origin, so the average energy becomes lowest for the case of $r^2$ potential.
These differences in the classical orbits are reflected in the behavior of the quantum-mechanical wave function. 
Namely, wave function with $l_z=0$ only has non-zero value at the origin.
The wave functions with $l_z=0$ suffers large deformation in the presence of the $1/r$ potential as we will see in the next section. 
Thus, the energy lowers quadratically as $\alpha$ increases.
On the other hand, those with $l_z \ne 0$ do not have large amplitude where the potential has large negative value, and suffers less deformation.
Thus, the energy lowers almost linearly with $\alpha$.

\begin{figure}[tb]
\begin{center}
\includegraphics[width=8cm]{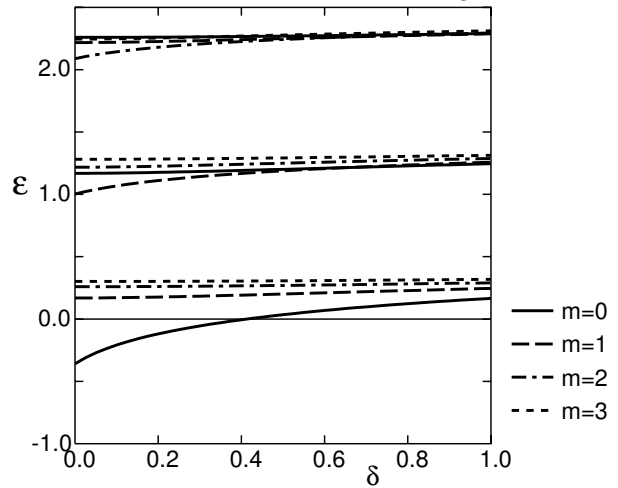}
\end{center}
\caption{Energy spectrum of states around an impurity with strength $\alpha=0.5$ as functions of $d$. 
The energy of the lowest states at $d=0$, the $l_z=n-m=0$ states, becomes higher as $d$ increases. At around $d \simeq \lb$, the lowest states become those with $m=0$.}
\label{fig3}
\end{figure}

We can expect that as the potential minimum at $r=0$ becomes shallower, the potential approaches the harmonic potential, and the energy of the $l_z=0$ state becomes higher than the $m=0$ state.
Thus, as the parameter $d$ in the potential increases the sequence of the energy changes.
The behavior of the energy spectrum at $\alpha=0.5$ as a function of $d$ is shown in Fig.~\ref{fig3}.
As expected, the level crossing occurs at around $d \simeq \lb$ except for the lowest Landau level, where state with $m=0$ and that with $l_z=0$ are the same state.
In this figure, we notice that only the $l_z=0$ states suffer large shift of energy as $d$ increases.
This behavior is understandable if we consider the classical orbits.
Namely, as $d$ increases only the potential around the origin changes.
Thus, only states which passes through a space around the origin suffers shift in the energy.

\section{Wave Functions}

\begin{figure}[tb]
\begin{center}
\includegraphics[width=7cm]{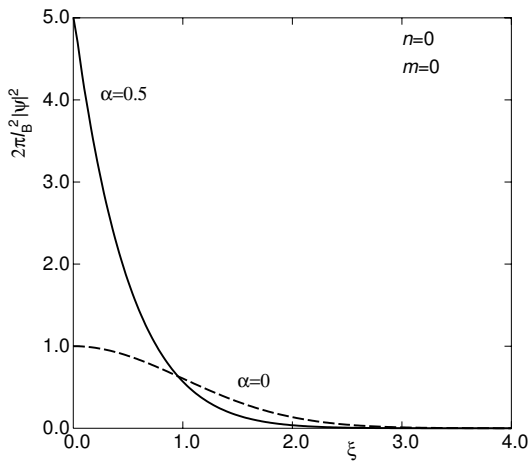}
\end{center}
\caption{Squared absolute value of the wave function with $n=m=0$,  $2\pi\lb^2|\psi_{0,0}(r,\theta)|^2$ at $\alpha=0.5$, solid line, and at $\alpha=0$, dashed line. }
\label{fig4}
\end{figure}

\begin{figure}[htb]
\begin{center}
\includegraphics[width=7cm]{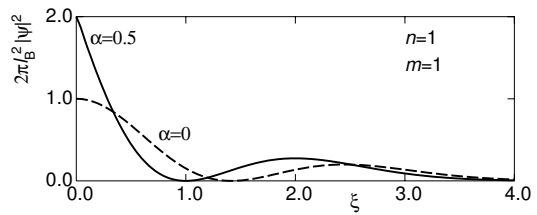}
\end{center}
\caption{Squared absolute value of the wave function with $n=m=1$,  $2\pi\lb^2|\psi_{1,1}(r,\theta)|^2$ at $\alpha=0.5$, solid line, and at $\alpha=0$, dashed line. }
\label{fig5}
\end{figure}

\begin{figure}[htb]
\begin{center}
\includegraphics[width=7cm]{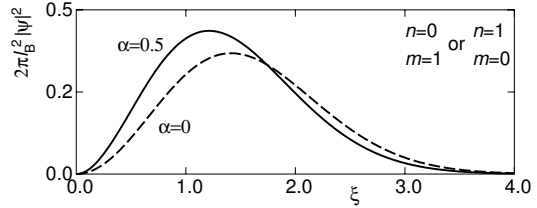}
\end{center}
\caption{Squared absolute value of the wave function with $n=0$ and $m=1$,  $2\pi\lb^2|\psi_{0,1}(r,\theta)|^2$ at $\alpha=0.5$, solid line, and at $\alpha=0$, dashed line. 
The wave function with $n=1$ and $m=0$ has the same form.
In this figure the vertical scale is enhanced by four times compared to Fig.~\ref{fig4} and \ref{fig5}.}
\label{fig6}
\end{figure}

In the absence of the impurity potential, the wave function is
\begin{widetext}
\begin{eqnarray}
\langle \vec{r} | n,m \rangle &\equiv&   \psi_{n,m}(r,\theta) \nonumber\\
&=& \frac{1}{\sqrt{2\pi}\lb}\sqrt{\frac{N!}{(N+|l_z|)!}}
\exp\left(-\frac{1}{4}\xi^2 +\I l_z\theta\right)
\left(\frac{1}{\sqrt{2}}\xi \right)^{|l_z|}L_N^{|l_z|}\left(\frac{1}{2}\xi^2\right) \,,
\label{eq:29}
\end{eqnarray}
\end{widetext}
where $\xi=r/\lb$, $l_z=n-m$, $N=n-(l_z+|l_z|)/2$, and polar coordinate $(r,\theta)$ is used.
At non zero $\alpha$, state with the same $l_z$ are mixed, and the wave functions are deformed from this form.
We will see in Figs.~\ref{fig4} to \ref{fig6} how these wave functions are deformed in the presence of the $1/r$ impurity potential.

In Fig.~\ref{fig4} the squared absolute values of the wave function $|\psi_{0,0}(r,\theta)|^2$ at $\alpha=0$ and that at $\alpha=0.5$ are shown.
The round Gaussian peak of the wave function at $r=0$ for $\alpha=0$ is raised and deformed to a cusp in the presence of the potential.
Actually, we can show analytically that the wave function with angular momentum zero has a cusp at $r=0$ by examining the Schr\"odinger equation at small $r$.
By considering the most important terms around $r=0$ we can show that
\be
|\psi_{n,n}(r,\theta)|^2 \simeq |\psi_{n,n}(0,0)|^2 \left(1-4\alpha\xi  \right) \,.
\ee

For other examples of the wave functions, we show in Fig.~\ref{fig5} the wave function at $n=1$ and $m=1$, and in Fig.~\ref{fig6} those at $n=0$ and $m=1$ or at $n=1$ and $m=0$.\cite{note2}
The former is another example of the wave function at $l_z=0$.
The latter is an example at $l_z=n-m= \pm 1 \ne 0$.
The deformation of the latter case is smaller than the $l_z=0$ case, since this state does not have amplitude at $r=0$ where the potential diverges.\cite{note1}
The smaller deformation is also reflected in the smaller energy lowering of these state at $l_z \ne 0$.


\section{Local Density of States}

The scanning tunneling spectroscopy (STS) measures local density of states as a function of energy $E$ and position $\vec{r}$.
We here examine how the states around the impurity potential show up in the STS experiment.
We consider an empty two-dimensional system with the Fermi energy $E_{\rm F}=0$.
Since, there is no electron in the two-dimensional system before electron tunnels into the system, interaction between electrons can be neglected, so single electron states considered so far give correct result.
Assuming a constant level broadening $\Gamma$ for states above the Fermi energy, we can write the local density of states (LDOS) as
\be
D(E,\vec{r}) = \frac{1}{\pi} \sum_n\sum_m\frac{\Gamma}{(E-E_{n,m})^2 + \Gamma^2} \, |\psi_{n,m}(\vec{r})|^2 \,,
\ee
where $E_{n,m}$ is the energy of the state with the quantum numbers $n$ and $m$.
This LDOS is proportional to the differential tunneling conductance, and measured in the experiment.

In the absence of the impurity potential, summation over $m$ makes this LDOS independent of the position, and energy dependence is given by the superposition of the Lorentzian distribution functions.
In the presence of the $1/\sqrt{r^2+d^2}$ potential at the origin, it becomes dependent on the distance from the origin.
We will investigate how these dependences reflect the potential form.

\begin{figure}[h]
\begin{center}
\includegraphics[width=7cm]{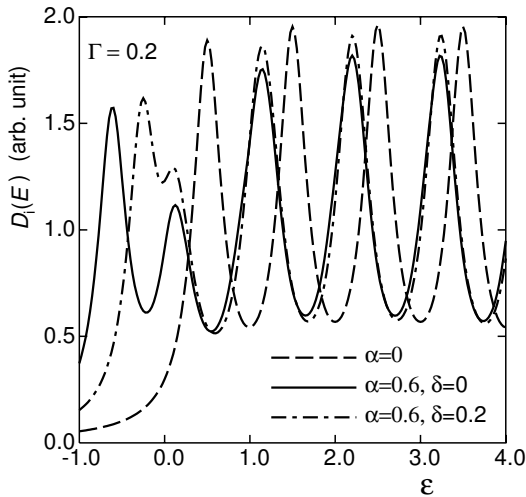}
\end{center}
\caption{Local density of states integrated in a circular area of radius $1.5\lb$ around the origin, where an impurity sits.
To make the results dimensionless $\hbar\oc D_{\rm i}(E)$ is plotted as functions of $\mathcal{E}$.
The dashed line shows the result for $\alpha=0$, the solid line shows that for $\alpha=0.6$, $\delta=0$, and the dash-dotted line shows that for $\alpha=0.6$, $\delta=0.2$.
Here the level broadening parameter $\Gamma=0.2\hbar\oc$ is used.}
\label{fig7}
\end{figure}

Having in mind comparing the result with existing experiments, we choose parameters $\alpha$ and $\Gamma$ to be $\alpha=0.6$ and $\Gamma=0.2\hbar\oc$.
As for the distance $d$ we consider two cases, $\delta=d/\lb=0$ and $\delta=0.2$.
First we show LDOS integrated in a circular area of radius $1.5\,\lb$ around the impurity in Fig.~\ref{fig7}:
\be
D_{\rm i}(E) = \int_0^{1.5\,\lb}\D r 2\pi r D(E,\vec{r}) \,.
\ee
The results for $\delta=0$ (solid line) and $\delta=0.2$ (dash-dotted line) almost coincides except for low energy region, where the energy of the $n=m=0$ state differs considerably between these two cases.
The result without the impurity potential ($\alpha=0$, dashed line) shows superposition of the Lorentzian distribution as stated above.
The shift of the peaks to the lower energy side at finite $\alpha$ indicates formation of bound states around the origin.
We will compare this result with experimental result later.

\begin{figure}[h]
\begin{center}
\includegraphics[width=7cm]{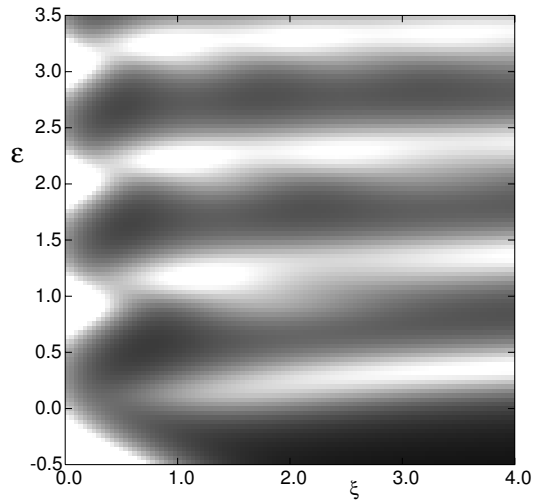}
\end{center}
\caption{Gray scale plot of the LDOS $D(E,\vec{r})$ for a system with impurity potential of $\alpha=0.6$ and $d=0$. The horizontal axis shows distance from the origin, the vertical axis shows energy in units of $\hbar\oc$, and the higher value of $D(E,\vec{r})$ is shown by brighter pixels.
Here the level broadening parameter $\Gamma=0.2\hbar\oc$.}
\label{fig8}
\end{figure}

\begin{figure}[h]
\begin{center}
\includegraphics[width=7cm]{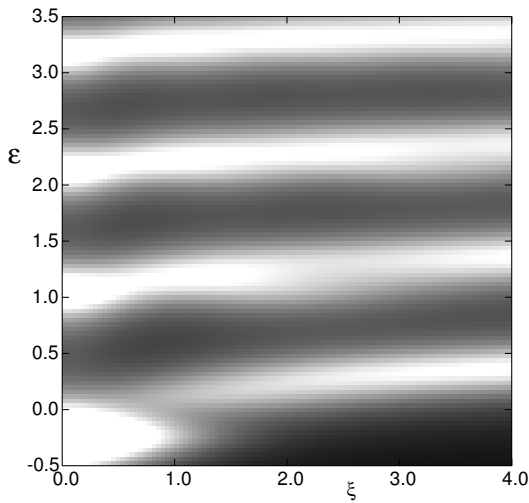}
\end{center}
\caption{Gray scale plot of the LDOS $D(E,\vec{r})$ for a system with impurity potential of $\alpha=0.6$ and $d/l=0.2$. The horizontal axis shows distance from the origin, the vertical axis shows energy in units of $\hbar\oc$, and the higher value of $D(E,\vec{r})$ is shown by brighter pixels.
Here the level broadening parameter $\Gamma=0.2\hbar\oc$.}
\label{fig9}
\end{figure}

To examine the details of the LDOS we plot the value of $D(E,\vec{r})$ in gray scale in Figs.~\ref{fig8} and \ref{fig9} for the case of $\delta=0$ and $\delta=0.2$, respectively.
In these figures, the horizontal axis shows distance from the origin, the vertical axis shows energy in units of $\hbar\oc$, and the higher value of $D(E,\vec{r})$ is shown by brighter pixels.
We notice that a series of ridges extend horizontally in these figures.
Each ridge is mainly composed of states of the same $n$.
The peak energies of the ridges tend to the energy of the Landau levels in the absence of the impurity $\mathcal{E} =n+1/2$ at $\xi \to \infty$.
A remarkable difference between the case of $\delta=0$ and $\delta=0.2$ is that the ridges are not continuous for the case of $\delta=0$, namely, several peaks are separated by saddle points along the ridges.

These peaks in Fig.~\ref{fig8} reflect the maximum of the squared wave functions.
Namely, for the case of states with Landau index $n=2$, the peak at $\xi=0$ and $\mathcal{E}=2$ reflects the fact that a state with $n=2$ and $l_z=0$ has energy of about $2\,\hbar\oc$ and the wave function is peaked at $\xi=0$.
The next peak at around $\xi=1$ and $\mathcal{E} \simeq 2.2$ comes from state with $n=2$ and $l_z=1$ and $l_z=-1$ whose wave functions are peaked around $\xi=1$.
For the case of $n=1$ series, similar correspondence can be seen between the peaks in LDOS and wave functions.

\begin{figure}[htb]
\begin{center}
\includegraphics[width=8.7cm]{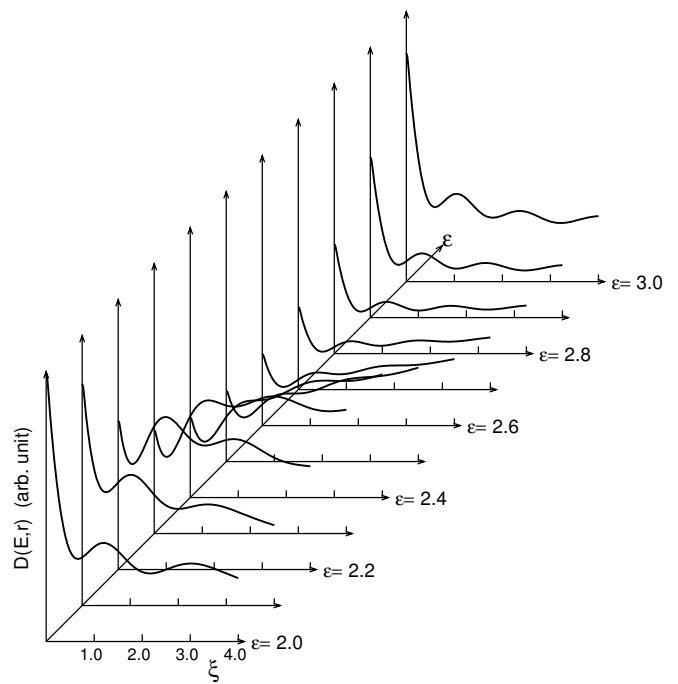}
\end{center}
\caption{Local density of states at fixed energies are shown as functions of $\xi$.
In this figure $\alpha=0.6$ and $\delta=0$.
Eleven values of $\mathcal{E}$ between $\mathcal{E}=2$ and $3$ with interval $0.1$ are chosen.
Here the level broadening parameter $\Gamma=0.2\hbar\oc$.}
\label{fig10}
\end{figure}

To give more quantitative picture of LDOS, we show in Fig.~\ref{fig10} the $\xi$-dependence of $D(E,\vec{r})$ at fixed energy $E$, at $\mathcal{E}=2$ , $2.1$ $\cdots$ and $3$.
Comparing these figures with the wave functions in Figs.~\ref{fig5} and \ref{fig6}, we can see that the saddle points in Fig.~\ref{fig8} are formed by the deformation of the wave function, and by the difference in energies of these states.
Figure 5 shows that the squared wave function of $l_z=0$ state decreases around $0.5 \le \xi \le 1.0$.
The increase of the squared wave function of $l_z \ne 0$ is not large enough to compensate this decrease, since the energy difference makes contribution to LDOS of these states at the energy of $l_z=0$ state smaller.

On the other hand, when $\delta=0.2$, these peaks of the wave functions overlap and saddle points are not so clear.
This is because the energies of the states are similar, and deformation of the wave function around the origin is smaller in this case.
Thus, even though the integrated LDOS looks similar for these two cases, the details of LDOS show clear difference.
When $\delta$ increases further, the ridges become more smooth and they approach horizontal ridges of system with $\alpha=0$.


\begin{figure}[htb]
\begin{center}
\includegraphics[width=8.7cm]{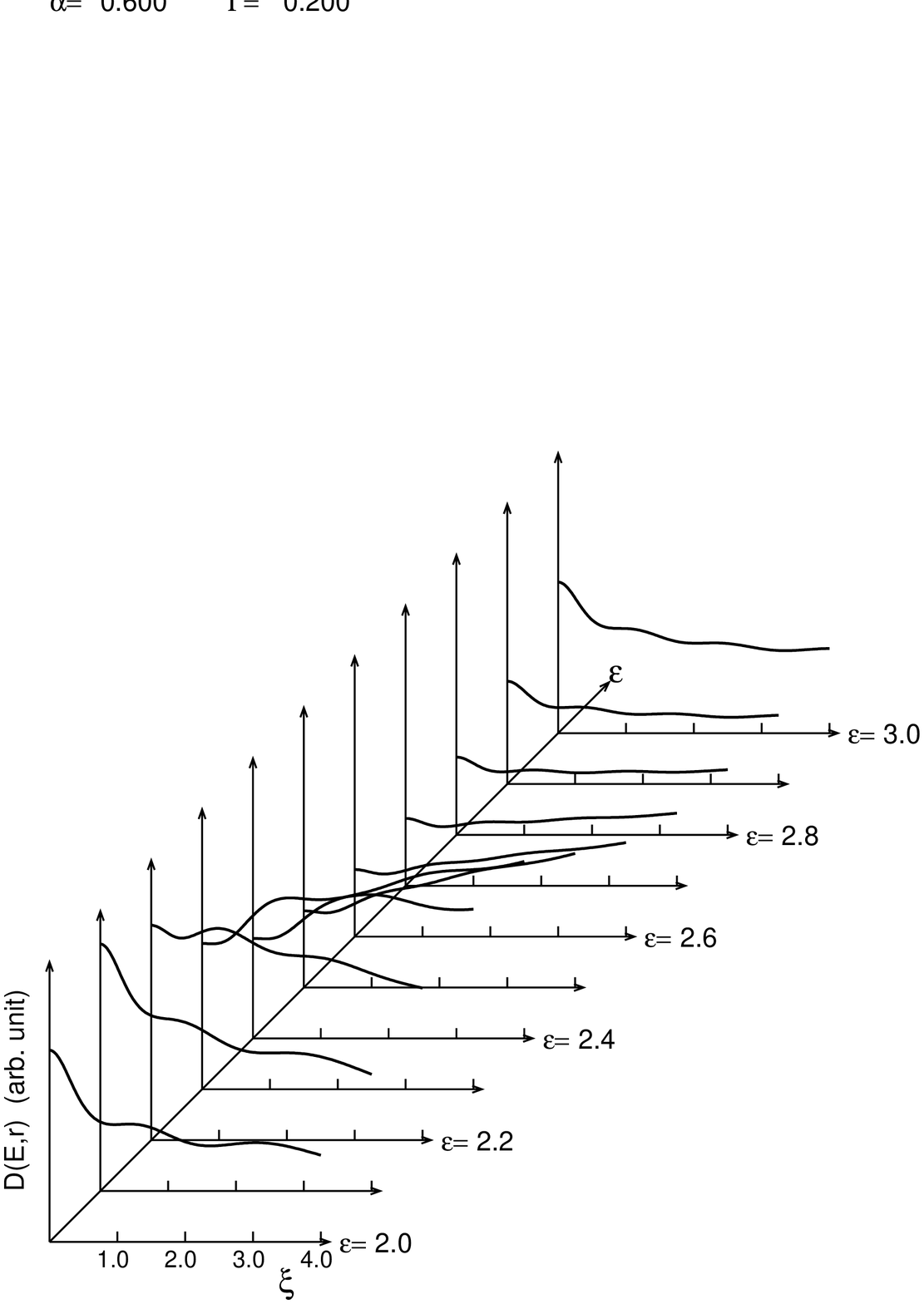}
\end{center}
\caption{Local density of states at fixed energies are shown as functions of $\xi$.
In this figure $\alpha=0.6$ and $\delta=0.2$.
Eleven values of $\mathcal{E}$ between $\mathcal{E}=2$ and $3$ with interval $0.1$ are chosen.
Here the level broadening parameter $\Gamma=0.2\hbar\oc$.}
\label{fig11}
\end{figure}

A similar figure of LDOS at given energies as Fig.~\ref{fig10} for the case of $\delta=0.2$ is shown in Fig.~\ref{fig11}.
In Figs.~\ref{fig10} and \ref{fig11}, we can also see how LDOS looks like at intermediate energies.
Especially, we observe that at the energy of the Landau level far from the impurity, $\mathcal{E}=2.5$, LDOS becomes smaller than surrounding regions.


\section{Discussion}

We have seen that the LDOS around an impurity potential has structure reflecting the strength and shape of the potential.
For the case of $d=0$, the peaks of LDOS have the same form as the wave function.
This theoretical result indicates that under suitable conditions part of the wave function is directly observed by the STS experiment.
In recent experiments on graphite, Niimi et al. experimentally observed LDOS in a strong magnetic field.\cite{niimi}
The observed structure of the LDOS between the Landau levels shows quite similar structure as the present result for pure Coulomb potential ($d=0$).

\begin{figure}[htb]
\begin{center}
\includegraphics[width=8cm]{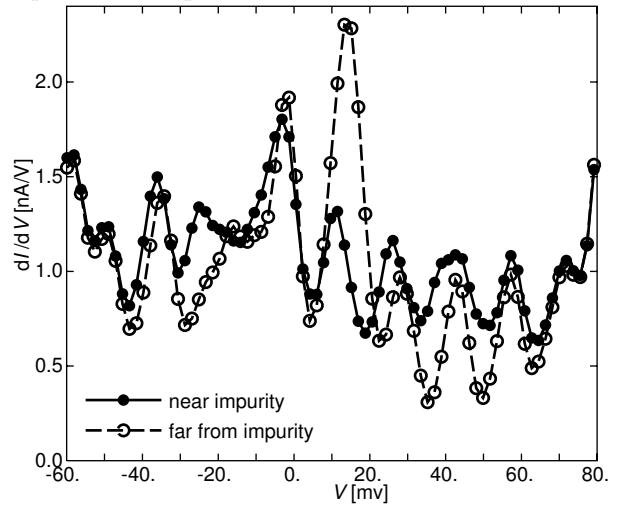}
\end{center}
\caption{Experimental differential tunnel conductance $\D I/\D V$ averaged over an area of 20$\times 20$ nm$^2$.
The data are taken at $B=6$\,T.
The horizontal axis shows bias voltage of the probe relative to the Fermi level.
The open circles show the data far from the impurity, and the closed circles show those around the impurity.
}
\label{fig12}
\end{figure}

Here we make quantitative comparison between the theory and their results.
First we show experimental differential tunnel conductance averaged over an area of 20$\times 20$ nm$^2$ in Fig.~\ref{fig12} to extract relevant parameters.
The horizontal axis shows bias voltage of the probe relative to the Fermi level.
This differential tunnel conductance is proportional to the LDOS, so $V>0$ part of this figure should be compared with Fig.~\ref{fig7}.
Far from the impurity, we see equally separated peaks corresponding to Landau levels.
The separation of peaks, about 14.6 mev, is consistent with the cyclotron energy $\hbar\oc$ calculated at $B=6$\,T and with electron effective mass of $0.057\,m_{\rm e}$.\cite{wallace, suematsu} 
The shape of the peaks looks like Lorentzian, and the parameter $\Gamma$ is deduced to be about 2\,mev.
The three-dimensionality of the graphite also should give width to the peaks, but the shape of the peak in that case is not Lorentzian, so we consider that the width comes from the life time of the states above the Fermi energy.

The peaks observed around the impurity are shifted to the low energy side.
From the size of the shift, we estimate approximate value for $\alpha$ to be 0.6.
If the impurity has a unit charge, $\alpha=0.6$ at $B=6$\,T is obtained by putting the dielectric constant $\epsilon\simeq 15\epsilon_0$.
Since graphite is a semimetal, the dielectric constant is not well defined.
Sometimes the value of $\epsilon=10\epsilon_0$ is used in the absence of the magnetic field as contribution from valence bands.
In the present situation, apart from contribution from valence bands, existing free carriers in the partially filled Landau level may contribute to screening of the impurity potential.
Virtual transitions between Landau levels also contribute to the dielectric constant.
The data in Fig.~\ref{fig12} show that the Fermi level ($V=0$) is a little above the hole Landau level peak at $V \simeq -2$\,mev.
Thus the Landau level is almost filled, and low density of free holes may be remaining.
These holes can screen negatively charged impurities, but are ineffective in screening the positively charged impurities.
Therefore, we can neglect the screening by free carriers in the present sample.
On the other hand, the contribution from the virtual transition depends on the wave vector $q$, and enhancement of $\epsilon$ at small $q$ is estimated by a simple calculation to be.\cite{alleiner}
\be
1+ 2q\lb \frac{\lb}{a_{\rm B}} \E^{-q^2\lb^2/2} \,,
\ee
where $a_{\rm B}$ is the Bohr radius.
In the present case the Bohr radius and the magnetic length is of the same order, and the enhancement factor at $q \simeq \lb^{-1}$ is about 2.
Thus the value $\epsilon=15\epsilon_0$ and $\alpha$ to be 0.6 are acceptable.

\begin{figure}[htb]
\begin{center}
\includegraphics[width=7cm]{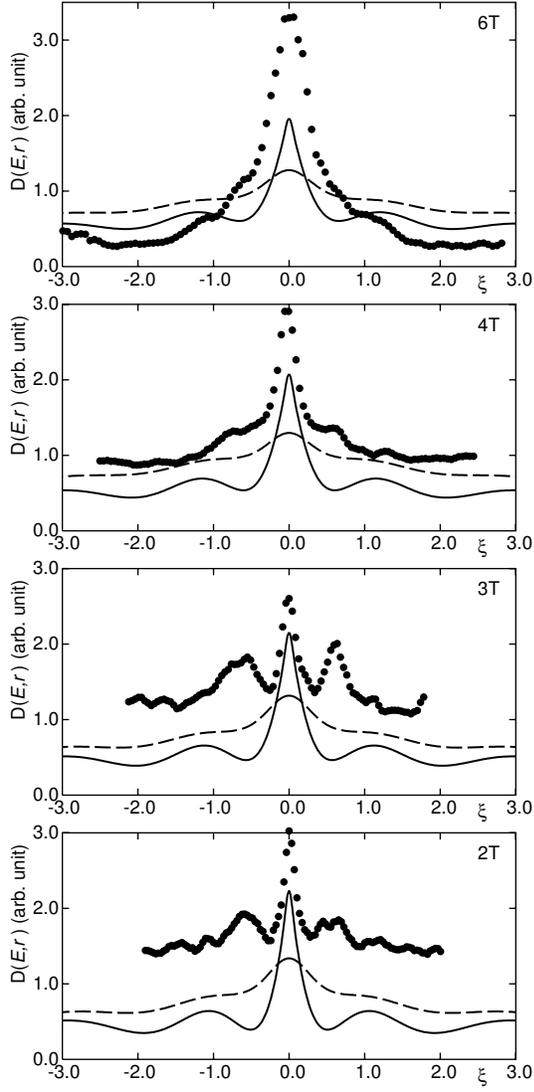}
\end{center}
\caption{
Experimentally observed differential tunnel conductance (closed circles) and theoretical LDOS for a potential with $\delta=0$ (solid line), and that for a potential with $\delta=0.2$ (dashed line) as functions of the distance from the impurity $\xi=r/\lb$.
The four values of magnetic fields, 2T, 3T, 4T and 6T are chosen.
}
\label{fig13}
\end{figure}

Using these parameters, we compare spatial dependence of the LDOS measured between the first and the second Landau levels.
In Fig.~\ref{fig13} experimental differential conductance and calculated LDOS are compared at several magnetic fields.
The value of $\alpha$ is chosen to be 0.6 at $B=6$\,T.
Since the $\alpha$ is inversely proportional to square root of the magnetic field, it increases as the magnetic field decreases.
The values are $\alpha=0.74$, 0.85 and 1.04 at $B=4$\,T, 3\,T and 2\,T, respectively.
The energy at which the theoretical LDOS is calculated is chosen as  $\mathcal{E}=2.0$, 1.9, 1.8 and 1.65 for $B=6$\,T, 4\,T, 3\,T and 2\,T, respectively, so that the central peak becomes highest.
Theoretical LDOS for potential with $d/\lb = 0.2$ are also plotted by dashed lines for comparison.
In this case, the values of $\mathcal{E}$ are 2.1, 2.05,1,95 and 1.85 for $B=6$\,T, 4\,T, 3\,T and 2\,T, respectively.
Comparison shows that the width of the central peak for $\delta=0$ is quite similar to the experimental width.
The sharpness of the experimental peak indicates that $d$ should be quite small; the potential at the impurity is well described by $1/r$ potential.

On the other hand, the position of the side peak, which forms ring-like structure around the central peak, is not well reproduced.
Experimental peaks are nearer to the origin.
If the value of $\alpha$ is much increased, the side peak comes nearer to the origin.
However, it will increase the binding energy considerably, and behavior of the averaged differential conductance cannot be reproduced.
Possible reason for the nearer side peak is the three dimensionality of the graphite.
We need further investigation for the explanation of this discrepancy.
In such investigation the actual band structure of the graphite should be taken into account.
Although this discrepancy remains, the qualitative agreement of the central peak shows that the potential shape can be inferred from the measurement of STS, and clarifies that the shape of the wave function can also be observed.

In this paper we investigated bound states around an impurity for an ordinary two-dimensional electron system, and compared the results with experimental data on graphite, which is a three-dimensional system, actually.
Recently true 2-d graphite is realized and called as graphene.
In this case the energy spectrum is not parabolic but linear.
Similar calculation for this case of linear spectrum is underway.
The result will be published in a near future.

\begin{acknowledgments}

This investigation is begun as collaboration with Prof. Fukuyama's group for the explanation of their experimental data.
Part of the results is included in ref.~2.
The author thanks Hiroshi Fukuyama and Yasuhiro Niimi for discussion and the use of their experimental data.
This work is partly supported by Grant-in-Aid for Scientific Research from the Japan Society for the Promotion of Science 18540311.

\end{acknowledgments}


\begin{thebibliography}{99}
\bibitem{note2} As noticed in \S 3, the squared absolute value of the wave function with $n=0$ and $m=1$ is the same as that with $n=1$ and $m=0$.
\bibitem{note1} Notice that the vertical scale of this figure is enhanced by four times compared to Figs~\ref{fig4} and \ref{fig5}.
\end{thebibliography}

\end{document}